# Putting the Context back into Memory

David A. Roberts

Micron Technology

*droberts@micron.com*

**Abstract**— Requests arriving at main memory are often different from what programmers can observe or estimate by using CPU-based monitoring. Hardware cache prefetching, memory request scheduling and interleaving cause a loss of observability that limits potential data movement and tiering optimizations. In response, memory-side telemetry hardware like page access heat map units (HMU) and page prefetchers were proposed to inform Operating Systems with accurate usage data. However, it is still hard to map memory activity to software program functions and objects because of the decoupled nature of host processors and memory devices. Valuable program context is stripped out from the memory bus, leaving only commands, addresses and data. Programmers have expert knowledge of future data accesses, priorities, and access to processor state, which could be useful hints for runtime memory device optimization. This paper makes context visible at memory devices by encoding any user-visible state as detectable packets in the memory read address stream, in a nondestructive manner without significant capacity overhead, drivers or special access privileges. We prototyped an end-to-end system with metadata injection that can be reliably detected and decoded from a memory address trace, either by a host processor, or a memory module. We illustrate a use case with precise code execution markers and object address range tracking. In the future, real time metadata decoding with near-memory computing (NMC) could provide customized telemetry and statistics to users, or act on application hints to perform functions like prioritizing requests, remapping data and reconfiguring devices.

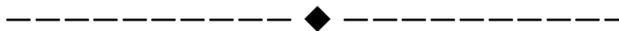

———————————— ◆ ————————————

## 1 INTRODUCTION

There are many use-cases for memory and storage request traces, including software optimization, hardware design and validation, system performance modeling and runtime optimization (e.g. page prefetching). Capturing memory traffic from production CPUs or GPUs is important because it gives insight into cache miss behavior and features such as metadata and coherency state updates in emerging protocols like CXL. Moreover, all memory traffic including prefetches are exposed, that may not be observable by software.

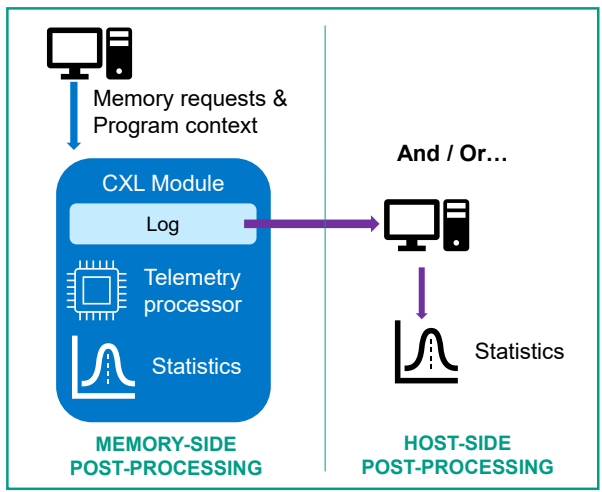

Fig 1. Emerging memory behavior analysis pipeline illustrating how a memory module could include programmable telemetry hardware.

Figure 1 illustrates a future "intelligent" configurable memory telemetry system, serving as a target architecture for the innovations in this paper. Current approaches to capturing main memory access traces include hardware simulation or recording with a protocol analyzer (e.g. for CXL). To try and map program activity to memory behavior, the beginning and end of regions of interest (ROIs) can be logged from a simulated system using special instructions recognized by a CPU simulator (e.g. gem5 [1]). Unfortunately, memory traces captured from real hardware (e.g. using a protocol analyzer) do not have a standard facility for recording ROIs. That makes aligning code events with memory trace segments very difficult. If this precise software-hardware context existed, memory traces could be analyzed to separately track statistics tied to program functions or even individual instructions. Examples include bandwidth, address locality, compressibility, page access frequencies (heat maps) to drive runtime memory tiering, and address sequences to drive page prefetching.

Hardware memory requests include cache-line reads (cache misses and pre-fetches) and writes. A useful capability would be to indicate which memory requests occur while a specific function (ROI) is executing, that could be of a very short duration (hundreds of nanoseconds). One approach to log the events would be to save them to a file, but there is no mechanism to accurately align them with timestamps in a memory trace. Another option could be to record events in a memory module (e.g. CXL) by writing to a designated I/O control register. There are several downsides to this, including the need for memory controller IP and a software driver.

Our technique avoids these issues by communicating metadata to a memory device in standard read requests with specially encoded addresses. Unlike writes, read requests do not modify data so they can overlay existing data, and they are not delayed by write-back caches. The new read requests conveying metadata need to be distinguishable from ordinary traffic. They must also be detectable at main memory in the presence of caching, reordering, prefetching, and data forwarding within a CPU. This paper describes novel address encoding and decoding algorithms to achieve these goals. Decoding can either be done in software or by a new hardware unit.

---

## 2 USE-CASES FOR CONTEXT AT MEMORY

The concept of communicating information to memory devices through read addresses was illustrated using ROI markers and memory object tracking. There are more potential use cases for embedding context hints in the memory stream. A user could program memory modules containing near-memory computing processors (NMC) with custom functions to filter, summarize and classify low-level statistics for end-to-end system optimization. For example, advising the programmer to reorder loops to improve access patterns on objects with poor data locality or moving them to faster memory tiers if they are heavily accessed. For runtime data prefetching (e.g. HoPP [2]), separate address sequences can be identified more easily within object boundaries, which are likely to exhibit consistent access patterns within their address ranges. Moreover, the programmer could convey information to a memory controller about how the data will be accessed, whether streaming, random access, or following a stride pattern. Those hints could allow hybrid memory modules with diverse memory technologies to select the best device, multi-level cell mode or caching policy to use for which data.

## 3 RELATED WORK

The HOPP system [2] filters and then logs address traces for accesses to frequently used pages, in a memory device. A Host processor reads the traces to run a page prefetching algorithm. Papers [3, 4, 5] include memory-side heat map telemetry units, which compute runtime statistics on observed memory requests. Our method is complimentary to memory-side telemetry systems because it can focus limited hardware tracking resources using contextual hints, for example, only counting or tracing accesses on certain objects.

## 4 MAILBOXES, PACKETS AND MESSAGES

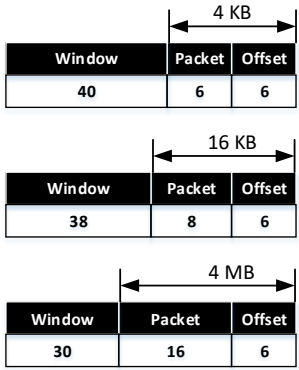

Fig. 2. Example host physical read address formats (each row is one cache line read request), trading off data payload size (encoded as a portion of the address) with increasing mailbox window sizes. Offset represents a 64B cache line size. For example, a 16-bit packet per read address requires a 4 MB naturally-aligned mailbox.

Figure 2 shows the address field of a memory read request. A 6-bit offset is implied due to 64-Byte cache lines typically being the smallest-request size coming from a CPU. We steal the next few bits to use as a data packet. The number of packet bits determines the address range (window) that needs to be monitored as part of a multi-packet transmission. The larger the window size the better, because a) packet bandwidth is greater and b) packets are scattered widely in the address space, reducing the likelihood of triggering prefetches.

Detecting metadata packets in a stream of unrelated read addresses requires multi-packet transmissions called messages. Messages originate as ordered sequences of data packets, plus a checksum packet (e.g. Cyclic Redundancy Check or CRC). The CRC must fit inside a single packet. The CRC may cover one or more data packets, but we found that 2 data packets (labeled A and B) per message is a good tradeoff between performance and reliability. The purpose of the CRC is so that a) we can identify which other packets are part of a message, and b) we know the original order of the packets, because the CRC check will only succeed with the correct data packet ordering. It was observed in practice that several repetitions of each message may be needed to ensure packets appear on the main memory bus, possibly due to forwarding logic within a processor that aborts redundant read requests (forwarding data from an earlier load to later ones).

The upper bits of a physical address, labeled Window in Figure 2, represent the range of addresses that packets from the same message could appear in. To reduce the amount of memory space that must be monitored to identify messages, a designated "mailbox window" can optionally be established when a program starts, known to both transmitter and receiver. Because reads do not modify data, the mailbox can be overlayed on top of any data owned by the application incurring no additional capacity overhead (Figure 3). Use of a dedicated mailbox window lets a decoder filter out most non-message read traffic (reads that must be decoded), making software-based decoding faster, but it is not required. Larger windows also reduce the impact of prefetching because it spreads out the addresses, reducing the likelihood of streams being detected and the prefetcher attempting to issue requests. Read prefetches could increase the likelihood of spurious packets appearing in the mailbox. To further reduce prefetches, a randomizer function [6] can be used to reduce correlation between packets and increase the distance between addresses, making it less likely to trigger a prefetch.

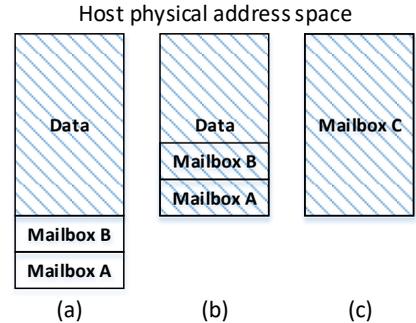

Fig. 3. Options for mailbox address range allocation. (a) Processes A and B allocate dedicated mailbox objects separate from program data, (b) mailboxes overlap data and (c) larger mailbox overlaps data.

Mailboxes need to be contiguous in the physical address space. This can be guaranteed by using an Operating System page size larger than the mailbox, or a memory allocator that maintains contiguous virtual and physical address range mappings. We use the latter method in our prototype.

## 5 ENCODING DATA USING READ ADDRESSES

When a program is started, it allocates a mailbox window that is naturally aligned in the physical address space. It then sends a "preamble" predefined sequence of packets in the mailbox range, that a receiver can look for in the memory trace. Once found, the receiver knows the (naturally aligned) physical (P) mailbox address so it can filter out any non-mailbox traffic. Then, some useful mailbox information is sent. This includes the virtual (V) address of the mailbox window (known only by the user process). The virtual-to-physical offset (V-P) can then be calculated at the receiver, allowing easy translation from physical trace addresses back to virtual addresses (e.g. objects of interest to the programmer). The Process ID can also be sent to support separate per-process mailboxes.

The user program uses a simple encoder library to send messages. To encode a message, "send" library function (Figure 5) is passed a message to inject into the trace. Alternatively, a dynamic binary translation tool (e.g. PIN [7]), or dynamically loaded library can intercept operations of interest like function calls and memory allocations, triggering message insertion. Long multi-message transmissions can embed a sequence number as part of their data payload if they need to be decoded in order. Then, each packet of the payload is assembled into an address, including the mailbox window base address (Figure 2). Finally, the cache line at that address is flushed (with writeback-invalidate instructions) then read, forcing a cache read miss.

## 6  MAILBOX DETECTION & PACKET DECODING

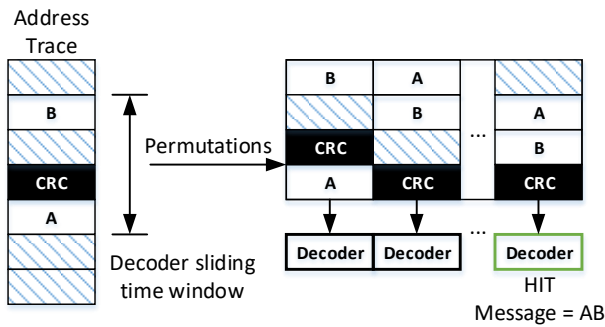

Fig. 4. Address trace message decoding procedure and hardware design

Figure 4 illustrates how messages are decoded, either in hardware or software. Decoding has two phases. The first is to identify the mailbox address, if any. If a mailbox is used, only addresses with a matching mailbox are considered.

The first phase locates the mailbox address range by scanning for the predefined preamble sequence, considering all addresses. The preamble sequence is a long, predefined repeated message, 100 packets for example. This preamble decoding step maintains a hash table of all observed read requests, indexed by mailbox-sized address ranges that could potentially be the real mailbox. This step can also detect the right CRC and mailbox size to use by maintaining a hash table for every supported window size, simultaneously trying to decode all sizes (using 8-bit packets with CRC8, 16-bit packets with CRC16, etc.) until the preamble is found. Once the specific mailbox address range is found, Phase 2 begins, and the decoder can ignore all addresses outside the mailbox range. This improves performance for software decoding because the decoder can ignore most memory requests.

In the second (running program) phase, the read addresses are scanned in time order and a sliding window of requests is decoded (we found that a window of 8 reads is reliable). A CRC check is performed on every possible permutation of 3 packets in the window. A successful check correctly identifies the transmitted packets and their original order. A hardware implementation for real-time decoding consists of a permutation network feeding a parallel array of CRC decoders.

## 7  IMPLEMENTATION & USE CASE DEMONSTRATION

The encoder was implemented as a C software library, and the decoder as a Perl program, although a real-time hardware implementation is feasible (see Figure 4). A test program was modified to include unique messages on entry and exit to functions. The initial proof-of-concept emulated a processor using the gem5 simulator [1] out-of-order detailed CPU and DRAM models, with several different types of prefetcher. This created realistic amounts of traffic to interfere with the new metadata messages. Main memory traces were saved to disk, then run through the decoder script, which successfully extracted the messages without packet loss or false positives when using 16-bit packets and CRCs (a window size of 4 MB according to Figure 2).

Once proven in simulation, the scheme was evaluated on hardware. Traces were recorded using an Experimental CXL-based Memory Request Logger, connected to a CXL-capable x86 CPU. The FPGA-based CXL card reserved 256 GB of DDR4 DRAM capacity for user applications, and another 256 GB to filter and log incoming CXL.mem request commands, timestamps and addresses with no slowdown. In the first experiment, the same test benchmark from simulation was run using a custom memory allocator library. It maintained a single contiguous virtual-to-physical address range, making it simpler to translate from physical (trace) to virtual (user code) by adding the fixed offset learned during mailbox detection. In practice, the reverse mapping could be done via a reverse page table in the operating system or in the memory module [2].

On hardware, the encoder and decoder software used in the simulated system functioned as expected. All function entry/exit messages were reliably extracted from their precise locations in the memory trace. In another hardware experiment, memory allocation (*malloc*) calls in a test benchmark were replaced with a wrapper function that encodes a message with a unique object ID, its virtual address and size. This enabled individual software objects to be tracked, and their trace requests isolated, over their lifetime.

```
template<typename T> void send_packet_addr( T
*pMailbox, T *message, uint64_t num_packets )
{
    // Encode message
    T *pData = message;
    int curr_pos = 0;

    while( curr_pos < num_packets )
    {
        // Assemble the packet read address
        char *pMeta = (char*)pMailbox +
(((uint64_t)*pData) << 6);

        // write message to memory
        flush( pMeta, 64 );
        volatile unsigned char temp = *((un-
signed char*)pMeta);
        flush( pMeta, 64 );

        pData++;
        curr_pos++;
    }

    // write checksum to memory
    T checksum = CRC( message, num_packets );
    char *pChecksum = (char*)pMailbox +
(((uint64_t)checksum) << 6);

    flush( (void*)pChecksum, 64 );
    volatile unsigned char temp = *pChecksum;
    flush( (void*)pChecksum, 64 );
}
```

Fig. 5: C code for sending a packet encoded as part of an address

```
pMailbox = mailbox_alloc();
MetaAddrMarker<CRCTYPE> marker1( "M1" );
…
for (int k = 0; k < NTIMES; k++)
{
    marker1.send( pMailbox );
    times[0][k] = omp_get_wtime();
    g->nbrscan();
    …
}
```

Fig.6. Neve benchmark main loop that runs for NTIMES iterations, with metadata mailbox allocation and loop marker function calls added.

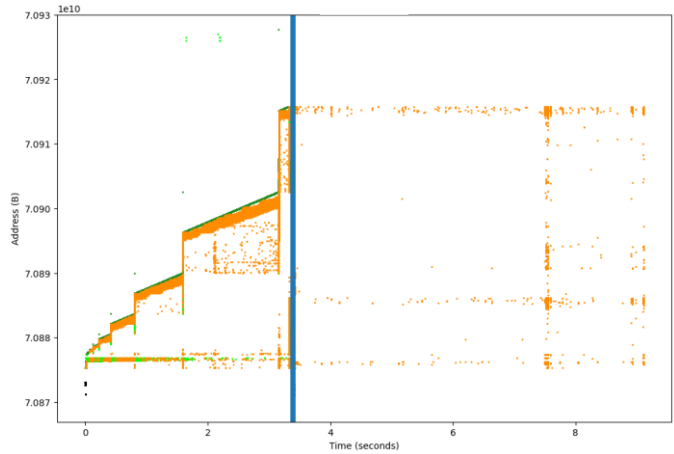

Fig. 7. Neve full trace with ROI marker overlay. Dark green dots are memory reads (CXL MemRd commands), light green are also reads (CXL MemRdData commands), orange are writes (CXL MemWr commands), and vertical blue lines are the markers for each loop iteration.

In the second experiment, the Neve [8] benchmark was modified to include a marker object using a specified CRC type and marker ID "M1". A marker "send" call was inserted in the main iteration loop (Figure 6). Every call to send embeds a marker and an incrementing call count into the address trace.

Figure 7 shows the resulting decoded memory address trace with loop iteration markers (blue lines) extracted from the trace. Figure 8 shows a zoomed in view of the same loop markers in Figure 7, complete with marker IDs and iteration numbers decoded and extracted solely from the CXL memory read address trace.

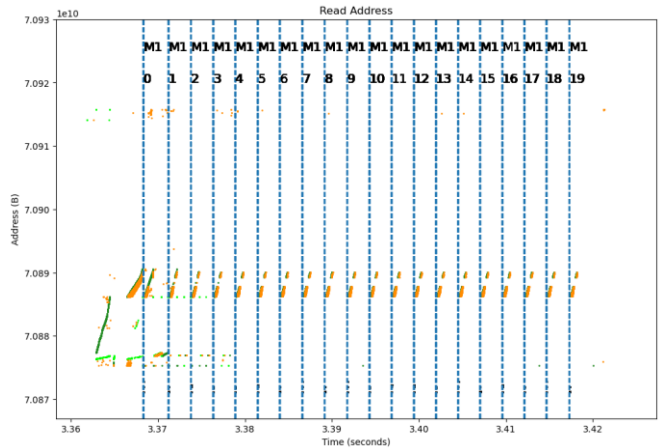

Fig. 8. Zoomed-in ROI showing loop iterations. All marker text and iteration numbers are precisely extracted from the memory trace.

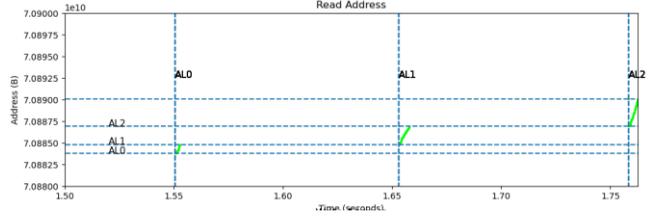

Fig. 9. Object tracking in a memory trace running the program in Figure 7.

In a third experiment, to demonstrate object tracking enabled by the new metadata technique, a simple test program was created. Three objects of increasing size were allocated then read in their entirety, with a delay between each one. Every *malloc* call sent metadata including a unique object ID number, and virtual address range belonging to the newly allocated object. While post-processing the hardware trace, the allocation timestamps were extracted and virtual addresses translated to physical addresses so they could be annotated directly onto the trace (Figure 9). By maintaining a map of the address space, the host processor or memory device can keep track of where each object of interest is located, isolating its associated memory traffic to provide programmer feedback or optimize runtime control mechanisms.

## 8 CONCLUSIONS

The proposed technique is a lightweight method of restoring valuable context to the memory request stream, because no special drivers or libraries are required as users simply access memory regions already allocated to them. This paper demonstrated use-cases enabling offline workload analysis and optimization from the perspective of the true main memory accesses, including hardware prefetch requests that were previously hidden from the programmer. Potential online use-cases were also discussed, enabling both automatic reconfiguration, and customized statistics feedback for end-to-end system optimization.

## ACKNOWLEDGMENT

This work was supported in part by the U.S. Department of Energy.